\documentclass[a4paper]{jpconf}

\usepackage{iopams}
\usepackage{amscd}
\usepackage{graphicx}
\usepackage{times}

\usepackage[breaklinks=true,colorlinks=true,linkcolor=blue,urlcolor=blue,
citecolor=blue]{hyperref}
\usepackage{epstopdf}
\usepackage{floatrow}
\usepackage{subfigure}
\usepackage{color}
\usepackage{xcolor}

\begin{document}

\title{Dynamical and quantum effects of collective dissipation in optomechanical systems}

\author{Albert Cabot}
\address{IFISC (UIB-CSIC), Instituto de Fisica Interdisciplinar y Sistemas 
Complejos, Palma de Mallorca, Spain.}
\ead{albertcabot@ifisc.uib-csic.es}

\author{Fernando Galve}
\address{IFISC (UIB-CSIC), Instituto de Fisica Interdisciplinar y Sistemas 
Complejos, Palma de Mallorca, Spain.}
\ead{fernando@ifisc.uib-csic.es}

\author{Roberta Zambrini}
\address{IFISC (UIB-CSIC), Instituto de Fisica Interdisciplinar y Sistemas 
Complejos, Palma de Mallorca, Spain.}
\ead{roberta@ifisc.uib-csic.es}

\begin{abstract}
Optomechanical devices have been cooled to ground-state and genuine quantum features, as well as long-predicted nonlinear phenomena, have been observed.
When packing close enough more than one optomechanical unit in the same substrate the 
question arises as to whether collective or independent dissipation channels are the correct description
of the system. Here we explore the effects arising when introducing dissipative couplings  between
mechanical degrees of freedom. We investigate synchronization, entanglement and cooling, finding that collective dissipation can drive 
synchronization  even in the absence of mechanical direct coupling, and allow to attain larger entanglement and optomechanical cooling. 
The mechanisms  responsible for these enhancements are explored and provide a full and consistent picture.  
\end{abstract}


\vspace{2pc}

\section{Introduction}

Cavity optomechanical systems (OMs) consist of a set of cavity light modes 
coupled to one or more mechanical elements typically by radiation pressure forces. 
OMs encompass many physical implementations ranging from the canonical 
Fabry-Perot resonator with a moving end mirror, to intracavity membranes, to the co-localized photonic and 
phononic modes of optomechanical crystals, to mention some \cite{1}. The nonlinear character of 
the optomechanical interaction enables OMs to work in various dynamical regimes 
and to exhibit a rich dynamical behavior. OMs can settle both in fixed points 
and display phenomena such as optical bistability \cite{2,3}, as well as 
into limit cycles or self-sustained oscillations \cite{4,5}, in which 
dynamical multistability is found \cite{6,7,8}. Moreover, arrays of 
coupled OMs can synchronize their self-sustained oscillations, as it has been 
shown both theoretically \cite{9,10,11} and 
experimentally \cite{12,13,14}. Spontaneous (or mutual) quantum synchronization (recently overviewed in Ref. \cite{55}) has also been considered in several systems 
\cite{55,30,31,lalo,extrasync1,extrasync2,extrasync4,extrasync5,extrasync6,crctn2,crctn3,crctn4}, also for CB dissipation \cite{30,31,lalo,extrasync2,extrasync4,extrasync5,extrasync6},
being optomechanical systems a promising platform to study this phenomenon
\cite{extrasync1,extrasync2,extrasync4,crctn2,crctn3,crctn4}. In the quantum regime optomechanical cooling allows efficient ground 
state cooling of the mechanical modes deep in the resolved sideband 
regime \cite{15,16}, and mechanical occupation numbers below one have been 
experimentally achieved \cite{17,18}. A variety of non-classical states and 
correlations have been predicted for OMs \cite{1}, some of them being recently 
observed \cite{19,20,21}. In particular entanglement in the asymptotic state has been considered 
between light and mirror \cite{22,23}, and  membranes \cite{24,25}, 
even at finite temperatures.  

The ubiquitous interaction of a system with its surroundings introduces noise 
and damping, and might lead to the complete erasure of quantum 
coherence \cite{26}. 
In the case of a spatially extended multipartite system -e.g. 
(optomechanical) array-, the spatial structure of the environment -like a field, a lattice or a 
photonic crystal- becomes crucial 
in determining different dissipation scenarios.
Two prototypical models are often considered: the separate 
bath model (SB) in which the units of the system dissipate into different 
uncorrelated environments, and the common bath model (CB) in which they 
dissipate collectively into the same environment. The particular dissipation 
scenario influences deeply the system. 
While SB dissipation usually destroys 
quantum correlations \cite{26},  dissipation in a CB leads to different results and enables phenomena such as 
decoherence free/noiseless subspaces \cite{27}, dark states \cite{zoller}, superradiance \cite{26,29}, dissipation-induced synchronization of linear networks of
quantum harmonic oscillators \cite{30,lalo} and of non-interacting spins \cite{extrasync5}, or no sudden-death of 
entanglement in systems of decoupled oscillators \cite{28,CB_ent}, that would be not present in these systems for SB dissipation. CB dissipation was 
first considered in the context of superradiance of atoms at distances smaller than the emitted 
radiation wavelength \cite{29} and 
often assumed to arise when the spatial extension of a system is smaller than the 
correlation length of the structured bath. A recent analysis of the CB/SB 
crossover in a lattice environment reveals the failure of this simple criteria and 
that actually collective dissipation can emerge 
even at large distances between the system's units (also for 2D and 3D environments \cite{31,CB_ent}). 

\begin{figure}[t!]
 \centering
 \includegraphics[width=0.7\columnwidth]{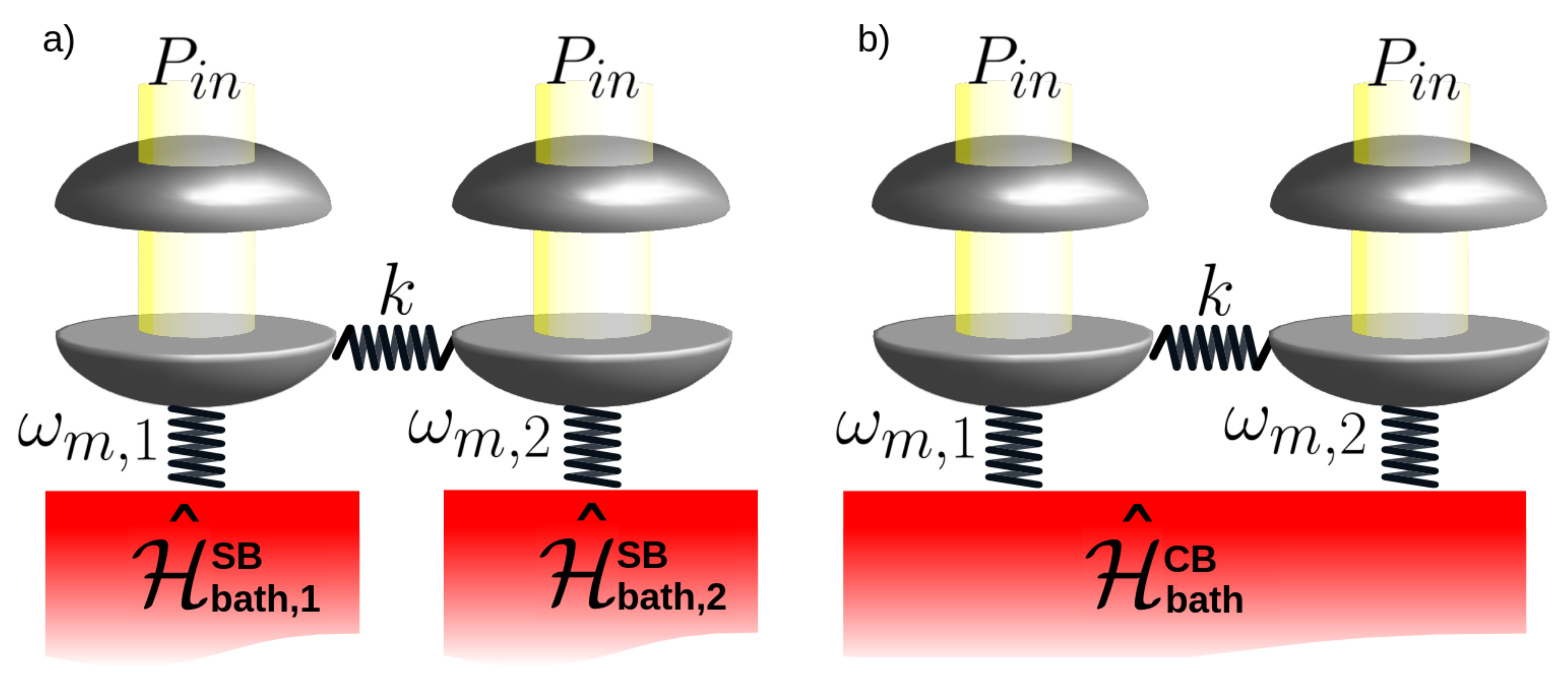}
 \caption{\label{figureone} In this work we consider an optomechanical system composed by two optical modes and two mechanical modes, and we allow for a finite
 mechanical coupling. We consider two dissipation scenarios. a) All the modes of the system dissipate into independent environments (separate baths). b) The mechanical modes 
 dissipate collectively into the same environment (common bath), while the optical modes dissipate into separate baths.}
\end{figure}

A detailed analysis of the 
CB/SB crossover in optomechanical arrays has not yet been reported and most of the works generally assume 
independent dissipation affecting optical and mechanical units. On the other hand, collective mechanical 
dissipation has been recently reported in some experimental platforms \cite{32,33},
such as OMs devices composed by two coupled 
nanobeams in a photonic crystal platform (environment). Photonic crystals are indeed also a known tool to suppress  mechanical dissipation \cite{34,35}. 
In these devices the main dissipation mechanism of the
nanobeams is found to be the emission of elastic radiation by the 
center of mass coordinate of the beams.   Furthermore,  an analogous  collective dissipation mechanism 
has been experimentally observed in piezoelectric resonators with a similar 
geometry \cite{36,37}. In general collective dissipation
can lead to quantitative and qualitative differences and we show how it can be 
beneficial  for collective phenomena and quantum correlations. In this work we investigate the effects of CB dissipation in optomechanical 
systems. We focus on a particular OMs consisting of two coupled mechanical elements 
each optomechanically coupled to a different optical mode (figure \ref{figureone}). We 
consider two dissipation schemes: independent  SB for all optical and mechanical modes (figure \ref{figureone}a), or CB for 
the mechanical modes and SB for the optical ones (figure \ref{figureone}b). We analyze the effects of collective dissipation both in the classical
and in the quantum regime, focusing on two phenomena already addressed in the literature in presence of SB, namely synchronization and entanglement  \cite{9,24}.
CB dissipation will be shown to have beneficial effects when compared to the SB case, both on classical synchronization (Sect. \ref{sect3}) and on asymptotic 
entanglement and optomechanical cooling (Sect. \ref{sect4}).

\section{Model and methods}
\label{sect2}

\subsection{Hamiltonian model}
 
A simple  configuration to assess collective or independent forms of dissipation consists of two optomechanical coupled units,
as sketched in Fig. \ref{figureone}. We allow for a finite mechanical-mechanical coupling to be compared with
dissipative coupling effects, while mixing between optical modes is not addressed here for simplicity. The total Hamiltonian is
\begin{equation}
\hat{\mathcal{H}}_S=\hat{\mathcal{H}}_1+\hat{\mathcal{H}}_2+\frac{k}{2}\big(\hat
{x}_1-\hat{x}_2\big)^2,
\end{equation}
where $\hat{\mathcal{H}}_{1,2}$ are the Hamiltonians for each optomechanical 
oscillator and the last term is the mechanical coupling between them. 
This kind of mechanical coupling can be found for instance in 
nanoresonators clamped to the same substrate as in Refs. \cite{32,33,cohcop}, as well as it can be induced optically using off-resonant lasers \cite{38}.
 Alternatively, interaction with a common (and single) optical field  can also induce coupling 
between  mechanical modes of OMs, as in Refs. \cite{11,14}.
The Hamiltonian for each unit of the OMs reads as \cite{40}:
\begin{equation} \label{eqHj}
\hat{\mathcal{H}}_j=\frac{1}{2}m\omega_{m,j}^2\hat{x}_j^2+\frac{1}{2m}\hat{p}
_j^2-\hbar\Delta_0\hat{a}^{\dagger}_j\hat{a}_j+i\hbar\sqrt{\frac{\kappa 
P_{in}}{\hbar\omega_{L}}}\big(\hat{a}^{\dagger}_j-\hat{a}_j\big)-\hbar\frac{
\omega_{c}}{L_{om}}\hat{x}_j\hat{a}^{\dagger}_j\hat{a}_j,\ \ j=1,2
\end{equation}
in the frame rotating with the input laser frequencies \cite{1}. The 
first two terms of the Hamiltonian correspond to the mechanical elements, 
described as harmonic oscillators of frequency $\omega_{m,j}$, effective mass 
$m$, and position and momentum operators $\hat{x}_j$, and $\hat{p}_j$, where 
$[\hat{x}_j,\hat{p}_j]=i\hbar$. The third and fourth terms describe the driven 
optical cavities in the frame rotating at the laser frequency, where $\Delta_0=\omega_L-\omega_c$, 
$\omega_c$ is the cavity resonance frequency, $\omega_L$ is the laser frequency, 
$P_{in}$ is the laser input power, $\kappa$ is the cavity energy decay rate, and 
$\hat{a}^{\dagger}_j$, $\hat{a}_j$ are the creation and annihilation operators of 
the light modes, with $[\hat{a}_j,\hat{a}^{\dagger}_j]=1$. 
 Notice that we make the simplifying assumption of 
identical optomechanical oscillators except for the mechanical frequencies, in 
which allowing disparity is essential for the study of synchronization. 
Finally the last term in Eq. (\ref{eqHj})
describes the optomechanical coupling with a strength parametrized by the 
effective optomechanical length $L_{om}$,  whose precise definition is system dependent and might involve the particular properties
of the interacting optical and mechanical modes (see for instance its definition for optomechanical crystals \cite{42}). Finally we remark that this model is appropriate for many different 
OMs as long as the parameters of the system, such as $m$ 
or $L_{om}$, appropriately refer to specific devices (like e.g. examples as diverse as in Refs. \cite{42,41}).

The optical and mechanical environments are described as an infinite collection 
of harmonic oscillators in thermal equilibrium. The standard model for the 
environment of an optical cavity, and the master/Langevin equations describing 
its dynamics, can be found in many textbooks on open quantum 
systems \cite{26,43,44}. We describe the mechanical environments 
using the model of Ref. \cite{45}, in which the Born-Markov master equations of 
two coupled $nonidentical$ harmonic oscillators for the CB/SB cases are derived,  
starting from the following Hamiltonians. 
In the SB case each mechanical oscillator is coupled to its own reservoir, 
so there are two different baths described by:
\begin{equation}
\hat{\mathcal{H}}^{SB}_{bath,j}=\sum^{\infty}_{\alpha=1} \hbar \, 
\omega_{\alpha,j}\hat{r}_{\alpha,j}^\dagger \hat{r}_{\alpha,j},\quad 
\hat{\mathcal{H}}^{SB}_{int,j}=\sum^{\infty}_{\alpha=1}  \lambda\, 
\hat{q}_{\alpha,j}\hat{x}_j, \quad j=1,2,
\end{equation}
where $\hat{r}_{\alpha,j}^\dagger$, $\hat{r}_{\alpha,j}$ are the creation and 
annihilation operators of the bath modes, and $\hat{q}_{\alpha,j}$ their 
position operators. On the other hand, in the CB case there is only one 
collective environment for both system units:
\begin{equation}\label{hamCB}
\hat{\mathcal{H}}^{CB}_{bath}=\sum^{\infty}_{\alpha=1} \hbar \, 
\omega_{\alpha}\hat{r}_{\alpha}^\dagger \hat{r}_{\alpha}, \quad 
\hat{\mathcal{H}}^{CB}_{int}=\sum^{\infty}_{\alpha=1} 2\, \lambda\, 
\hat{q}_{\alpha}\hat{x}_+,
\end{equation}
resulting in coupling through the center of mass coordinate  $\hat{x}_+=(\hat{x}_1+\hat{x}_2)/\sqrt{2}$ \cite{30,31,26,27,29,28,CB_ent,32}.
Notice that we have included a factor two in Eq. (\ref{hamCB}) to enforce an equal energy damping rate for both models: 
in SB case two independent channels dissipate, in CB case only one coordinate (center of mass) dissipates, but it is coupled
to the bath with double strength $2\lambda$. This allows for a quantitative, not only qualitative, comparison of both dissipation regimes.

\subsection{Dimensionless parameters and observables}
\label{sect2.2}
 A set of 
dimensionless observables and parameters can be introduced similarly to 
Ref. \cite{6}. First we define the quantities 
$x_{om}=\kappa L_{om}/\omega_c$ and $n_{max}=4 
P_{in}/\hbar\omega_L\kappa$, which we use to achieve a dimensionless time and dimensionless operators of the system:

\begin{equation}
t'=\kappa t,\quad \hat{x}_j'=\frac{\hat{x}_j}{x_{om}},\quad 
\hat{p}_j'=\frac{\hat{p}_j}{m\,\kappa\, x_{om}}, \quad 
\hat{a}_j'=\frac{\hat{a}_j}{\sqrt{n_{max}}}, 
\end{equation}
where the dimensionless quantities are  denoted by a prime. 
Then the following dimensionless parameters appear in the equations of motion: 

\begin{equation}
\omega_{m,j}'=\frac{\omega_{m,j}}{\kappa},\ \ \Gamma'=\frac{\gamma}{\kappa},\ \ K_c'= 
\frac{k}{m\kappa^2},\ \ \Delta_0'=\frac{\Delta_0}{\kappa},\ \ \mathcal{P}'=\frac{4P_{in}\omega_c}{mL_{om}^2\kappa^4}.
\end{equation}

 Unless stated otherwise we will work with these dimensionless parameters in the following, and thus we can drop the primes.

\subsection{Nonlinear classical equations}

After defining the set of dimensionless parameters and observables used in this work, we can write down the dimensionless equations describing the classical
dynamics of this system \cite{9}, both for SB and CB.
The classical dynamics  can be obtained by the quantum equations of motion taking the first moments
and making the approximation of factorizing expectation values of nonlinear terms, as usual. The quantum dynamics 
can be obtained either from a master equation approach as well as from the quantum Langevin equations of the system (see next section) \cite{1,44,45}:
\begin{equation}\label{cl1}
\dot{a}_j=\bigg[i(\Delta_0+x_j)-\frac{1}{2} \bigg]a_j+\frac{1}{2},\quad j=1,2,
\end{equation}
\begin{equation}\label{cl2}
\dot{x}_j=p_j,\quad 
\dot{p}_j=-\omega^2_{m,j}x_j+(-1)^{j}K_c(x_1-x_2)-\mathcal{D}_{SB/CB}+\mathcal{P
}|a_j|^2,
\end{equation}
where  variables without hat denote expectation values, e.g., $x=\langle\hat{x}\rangle$. The mechanical dissipation terms are defined as 
$\mathcal{D}_{SB}=\Gamma p_j$ for the separate bath case \cite{9}, and 
$\mathcal{D}_{CB}=\Gamma(p_1+p_2)$ for the common bath case \cite{45}.

The dynamical behavior of optomechanical systems can be very diverse depending on the  parameter choice \cite{1}. When the mechanical resonator is 'too slow' with 
respect to the cavity lifetime (low mechanical quality factor, and $\omega_m$ smaller than $\kappa$, in dimensional units), the intracavity power follows adiabatically
changes in the mechanical displacement, and a fixed point in which radiation pressure equilibrates with the mechanical restoring force is reached \cite{2}.
Otherwise, when the mechanical resonator is able to follow the fast cavity dynamics (high quality factors, and $\omega_m$ comparable to $\kappa$),
dynamical backaction
enables striking phenomena such as damping (cooling) or anti-damping (heating) of the mechanical motion by the optical cavity \cite{1,46}. In fact 
optomechanical cooling is enhanced in the red-sideband regime ($\Delta_0<0$) whereas anti-damping in the blue-sideband regime ($\Delta_0>0$). For enough 
laser power, $\mathcal{P}$, anti-damping can overcome intrinsic mechanical damping resulting in a dynamical instability and self-sustained 
oscillations \cite{1,46}. In these conditions the optomechanical system experiences a supercritical Hopf bifurcation when varying continuously $\Delta_0$
from negative to positive values \cite{6,47}. Moreover, for strong driving a dynamical multistability is found,
and self-sustained oscillations are possible even in the red-sideband regime \cite{6}.
The regime of self-sustained oscillations has been explored in the context of spontaneous synchronization 
in presence of independent dissipation for two coupled mechanical units \cite{9}
and in Section \ref{sect3} we are going to address the effect of dissipative coupling (CB).

\subsection{Linear quantum Langevin equations}

From the above Hamiltonian model and the input-output formalism we can obtain a 
set of Langevin equations describing the noise driven damped dynamics of the 
operators \cite{1} (see also the \ref{appA}), that will be used in Sect. \ref{sect4} to explore 
a possible improvement of non-classical effects in presence of collective dissipation.
We are going to consider the system in a stable and stationary state where 
a linearized treatment is a suitable approximation for the fluctuations dynamics
(for more details see for instance Refs. \cite{1,22,48}).
Low temperatures and high mechanical and optical quality 
factors \cite{1} also contribute to maintain low levels of noise (fluctuations). Indeed setting the OMs in a stable fixed point \cite{22,48} the fluctuation 
operators can be defined as:
$\delta\hat{O}= \langle{O}\rangle_{st}-\hat{O}$, where 
$ \langle{O}\rangle_{st}$ is the constant solution of equations (\ref{cl1}) and (\ref{cl2}), and the linear 
equations for the fluctuation operators are:

\begin{equation}\label{lan1}
\delta\dot{\hat{Q}}_j=-(\langle{x}_{j}\rangle_{st}+\Delta_0) \delta 
\hat{P}_j-\langle{P}_{j}\rangle_{st}\, \delta \hat{x}_{j}-\frac{1}{2} \delta \hat{Q}_j+ 
\sqrt{\frac{\omega_{m}}{\mathcal{P}}}\,\hat{Q}_{in,j}, \quad j=1,2,
\end{equation}

\begin{equation}\label{lan2}
\delta\dot{\hat{P}}_j=(\langle{x}_{j}\rangle_{st}+\Delta_0) \delta \hat{Q}_j+\langle{Q}_{j}\rangle_{st}\,
\delta \hat{x}_j-\frac{1}{2} \delta \hat{P}_j+ 
\sqrt{\frac{\omega_{m}}{\mathcal{P}}}\,\hat{P}_{in,j},
\end{equation}

\begin{equation}\label{lan3}
\delta\dot{\hat{x}}_j=\delta \hat{p}_j-d^x_{SB/CB}+ \hat{x}_{in,j},
\end{equation}

\begin{equation}\label{lan4}
\delta\dot{\hat{p}}_j=-\omega^2_{m}  \delta \hat{x}_j+(-1)^j K_c\big(\delta 
\hat{x}_{1}-\delta \hat{x}_{2}\big)-d^p_{SB/CB}+\mathcal{P} 
(\langle{Q}_{j}\rangle_{st}\,\delta \hat{Q}_j+\langle{P}_{j}\rangle_{st}\,\delta \hat{P}_j)+\omega_{m} \, 
\hat{p}_{in,j}.
\end{equation}
In the following we are going to consider the light quadratures 
$\hat{Q}_j=(\hat{a}_j+\hat{a}_j^{\dagger})/\sqrt{2}$ and  
$\hat{P}_j=i(\hat{a}_j^{\dagger}-\hat{a}_j)/\sqrt{2}$. We also notice 
 that we are considering identical optomechanical oscillators in the study of quantum correlations (Sect. \ref{sect4}). 
As usual, the fluctuation operators are rescaled by the parameter 
$\zeta=\tilde{x}_{zpf}/x_{fwhm}$ with $\tilde{x}_{zpf}=\sqrt{\hbar/m\omega_m}$, 
so that $\delta \hat{O}_{new}=\delta \hat{O}_{old}/\zeta$. 

The mechanical dissipation terms are:
\begin{equation}
\label{dissCBSB}
d^{x}_{SB}=\Gamma\delta\hat{x}_j, \quad d^p_{SB}=\Gamma\delta\hat{p}_j,\quad 
d^{x}_{CB}=\Gamma\big(\delta\hat{x}_1+\delta\hat{x}_2\big),\quad 
d^p_{CB}=\Gamma\big(\delta\hat{p}_{1}+\delta\hat{p}_{2}\big),
\end{equation}
where the known equivalence in position and momentum damping results from the 
rotating wave approximation in the system-bath coupling, a valid approximation 
in the
limit $\Gamma\ll\omega_m$ (i.e. high mechanical quality 
factors) \cite{26}. In the Markovian limit, the zero mean Gaussian noise terms of equations (\ref{lan1}) to (\ref{lan4}) are characterized by the following  symmetrized
correlations \cite{22}:

\begin{eqnarray}\label{corr}
\langle \big\{\hat{Q}_{in,j}(t),\hat{Q}_{in,j}(t')\big\}\rangle&=&\langle 
\big\{\hat{P}_{in,j}(t),\hat{P}_{in,j}(t')\big\}\rangle=\delta(t-t'), \quad 
j=1,2,\\
\frac{1}{2}\langle 
\big\{\hat{x}_{in,j}(t),\hat{x}_{in,j}(t')\big\}\rangle&=&\frac{1}{2}\langle 
\big\{\hat{p}_{in,j}(t),\hat{p}_{in,j}(t')\big\}\rangle=\Gamma(2n_{th}
+1)\delta(t-t'),
\end{eqnarray}
where $\{...\, ,\,...\}$ denotes anticommutator, $\langle ... \rangle$ 
denotes an ensemble average, $\delta(t-t')$ is the Dirac delta, and 
$n_{th}=(\exp[\hbar\omega_m/k_B T]-1)^{-1}$ is the phonon occupancy number of the 
mechanical baths, assumed to be at the same temperature $T$.  The light noise correlations for the corresponding quadratures are obtained
by considering optical environments at zero temperature for which the only nonvanishing correlations are $\langle \hat{a}_{in,j}(t)
\hat{a}^\dagger_{in,j}(t') \rangle=\delta(t-t')$ for $j=1,2$ \cite{22}. In the CB case 
additional cross-correlations appear:
\begin{equation}\label{crossnoise}
\frac{1}{2}\langle 
\big\{\hat{x}_{in,1}(t),\hat{x}_{in,2}(t')\big\}\rangle=\frac{1}{2}\langle 
\big\{\hat{p}_{in,1}(t),\hat{p}_{in,2}(t')\big\}\rangle=\Gamma(2n_{th}
+1)\delta(t-t').
\end{equation}

Finally we remark that as the fluctuations dynamics is linear and the noise is 
Gaussian, states initially Gaussian   will remain Gaussian at all 
times \cite{50,51}. Notably Gaussian states are completely characterized by 
their first and second moments, which are all encoded in the covariance matrix 
of the system \cite{50,51}. Rewriting the Langevin equations (\ref{lan1}-\ref{lan4}) as 
$\dot{\vec{R}}=\mathcal{M}\vec{R}+\vec{D}$, where $\vec{R}$ is the vector of the 
position, momentum, and light quadratures of the system, $\vec{D}$ is the vector 
containing the noise terms, and $\mathcal{M}$ is the matrix generating the 
dynamics of the system, the following equation for the covariance matrix can be 
written down:
\begin{equation}\label{coveq}
\dot{\mathcal{C}}=\mathcal{M}\, \mathcal{C}+\mathcal{C}\, 
\mathcal{M}^T+\mathcal{N},
\end{equation}
with the covariance matrix  and the noise covariance matrix defined respectively 
as:
\begin{equation}
\mathcal{C}_{ij}(t)=\frac{\langle R_i(t) R_j(t)+ R_j(t) R_i(t)\rangle}{2}, \quad 
\mathcal{N}_{ij}(t)=\frac{\langle D_i(t) D_j(t)+ D_j(t) D_i(t)\rangle}{2},
\end{equation} 
with $i,j=1,...,8$. By solving equation (\ref{coveq}) at any time we can completely 
characterize the quantum correlations present in the system \cite{52,53}.

\section{Classical synchronization}
\label{sect3}

In this section we analyze the effects of collective dissipation on the 
synchronization of the first moments of the system, or classical 
synchronization. In the SB case classical synchronization has been already 
studied in Ref. \cite{9}, where it has been shown that the OMs described by 
equations (\ref{cl1}) and (\ref{cl2}) can synchronize with a locked phase difference either 
tending to 0, in-phase synchronization, or to $\pi$, anti-phase synchronization. 
To study the presence of synchronization we integrate numerically equations (\ref{cl1}) 
and (\ref{cl2}) in both CB/SB cases and in a parameter region in which the system is 
self-oscillating. Once the system is in the steady state we compute a 
synchronization measure. Synchronization  is characterized by a Pearson 
correlation function defined as:
\begin{equation}\label{sync}
C_{x_1,x_2}(t,\Delta t)=\frac{\overline{\delta x_1\delta 
x_2}}{\sqrt{\overline{\delta x_1 ^2}\, \, \overline{\delta x_2^2}}},
\end{equation}  
where the bar represents a time average with time window $\Delta t$, i.e. 
$\overline{x}=\frac{1}{\Delta t}\int_t^{t+\Delta t}ds\, x(s)$, and $\delta 
x= x(t)-\overline{x}$. This correlation function has proven to be a useful 
indicator of synchronization both in classical and quantum 
systems \cite{55,54}. In particular this measure provides an absolute scale for synchronization strength as it
takes values between -1 
and 1, where -1 means perfect anti-phase synchronization and 1 perfect in-phase 
synchronization. 
Furthermore this measure can be generalized also in presence of time delays
accounting for regimes where 
OMs synchronize with a phase difference depending on the 
parameter values. In this case two delayed functions, $x_1(t)$ and  $x_2(t+\tau)$, need to be considered
in  Eq. (\ref{sync}). The best figure of merit accounting for delay is found maximizing the synchronization measure
for different values of $\tau$: in particular we take an 
interval of time $\Delta t$ such that it includes many periodic oscillations, 
and we compute the correlation function (\ref{sync}) between $x_1(t)$ and $x_2(t+\tau)$, 
where $\tau$ is varied between $[0,\mathcal{T}]$, being $\mathcal{T}$ the period 
of the oscillation. By keeping the maximum correlation $C_{max}$, and the 
temporal shift at which it occurs $\tau_{max}$, we obtain a measure of the 
degree or quality of synchronization, $C_{max}$, and the locked phase 
difference, $\tau_{max}/\mathcal{T}$, in units of $2\pi$. 

\begin{figure}[H]
 \centering
 \includegraphics[width=0.85\columnwidth]{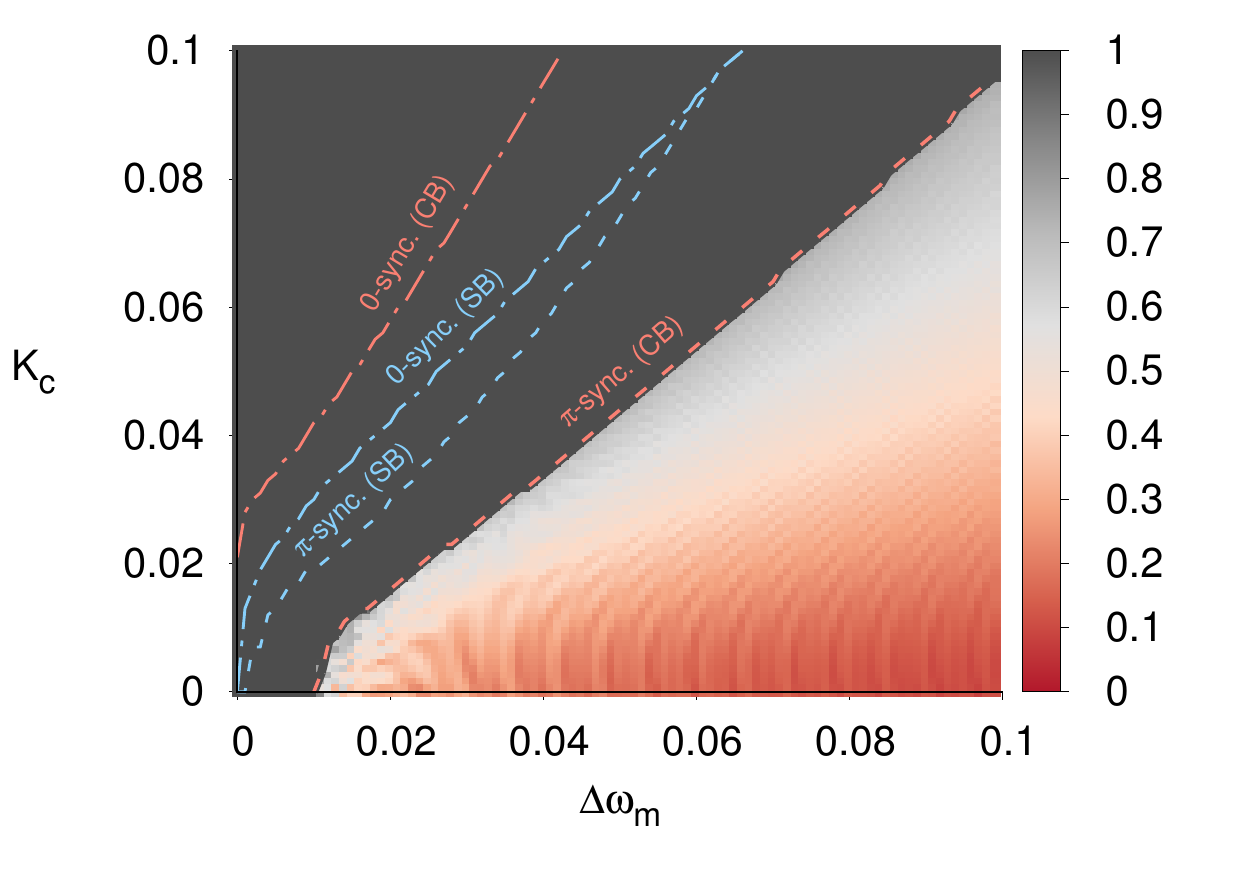}
 \caption{\label{figuretwo} Colorscale: degree of synchronization $C_{max}$
 between the 
 average mechanical positions of two 
coupled OMs  for the CB case, as a function of the mechanical 
frequency detunig, $\Delta\omega_m$, and the mechanical coupling, $K_c$. The 
fixed parameters take the following values: $\mathcal{P}=0.36$, $\Delta_0=1$, 
$\Gamma=0.01$, and $\omega_{m,1}=1$ (like in 
Ref. \cite{9}). Dashed lines indicate the onset of 
anti-phase synchronization. Dashed-dotted lines indicate the onset of in-phase 
synchronization. Salmon lines correspond to the CB case, and blue lines to 
the SB case.}
\end{figure}

The degree of synchronization accounting for delay, $C_{max}$, 
as a function of the frequency detuning, 
$\Delta\omega_m=\omega_{m,2}-\omega_{m,1}$, and the mechanical coupling 
strength, $K_c$, is shown in figure \ref{figuretwo}, where synchronization is found for coupling overcoming detuning
between the mechanical oscillators.
The case of dissipation in a CB is represented also including a comparison  with the SB case studied in 
Ref. \cite{9} in which a similar synchronization diagram was already observed. 
In the synchronized regions, $C_{max}\approx1$, the 
mechanical elements oscillate at the same frequency describing sinusoidal 
trajectories with constant amplitude. Conversely, when there is no 
synchronization ($0 \lesssim C_{max}\lesssim 0.6$ in the represented case), the mechanical oscillations have 
different frequencies and display strong amplitude modulations. 

In presence of synchronization two further regimes are recognized depending on whether the locked 
phase difference tends to $0$ or to $\pi$. At threshold the system first anti-synchronize and  
by further  increasing the coupling for a given detuning there is a transition to in-phase synchronization
(respectively, dashed and dashed-dotted 
lines of figure \ref{figuretwo}, for both CB (salmon lines) and SB (blue lines)). 
The comparison allows to identify how collective dissipation favors spontaneous synchronization: indeed, the
synchronized area for CB is much larger than for SB, as the first anti-phase synchronization threshold coupling is lowered (salmon and blue dashed `$\pi$-sync.' lines).
The anti-phase synchronization
regime is also enlarged for CB case as the second threshold to in-phase synchronization increases for CB with respect to SB.
The fact the anti-phase synchronization is favored is consistent with the particular dissipation form in the CB case: it is the 
common coordinate $x_1+x_2$ that is damped by the environment, see Eq. (\ref{cl2}), leading
to a `preferred' amplitude of motion in $x_1-x_2$, and thus an anti-phase locking.

\begin{figure}[H]
 \centering
 \includegraphics[width=1\columnwidth]{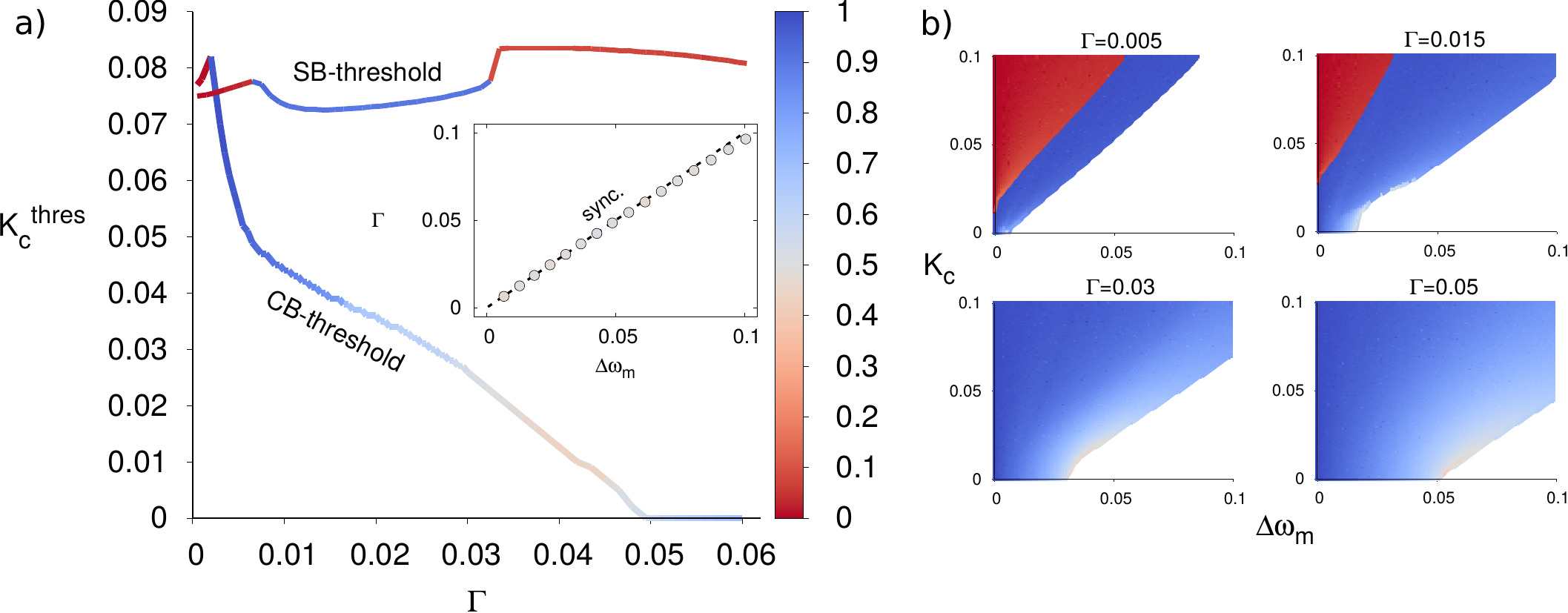}
 \caption{\label{figurethree}   a) Minimum mechanical coupling 
necessary to synchronize the oscillators, $K_c^{thres}$, as a function of the 
dissipation rate, $\Gamma$. The other parameters are fixed to: $\mathcal{P}=0.36$, 
$\Delta_0=1$, $\omega_{m,1}=1$, and $\Delta\omega_m=0.05$. Colorscale: locked phase difference in units of $\pi$. Inset: minimum dissipation rate, $\Gamma$,
necessary to synchronize the oscillators  in the CB case, for a given 
detuning, $\Delta\omega_m$, with $K_c=0$, and the other parameters as in the main panel. The dashed black line corresponds to $\Gamma=\Delta\omega_m$.
 b) Panels: locked phase difference (using the same colorscale as in Fig. \ref{figurethree}a) as a function of the mechanical frequency detunig, 
$\Delta\omega_m$, and the mechanical coupling, $K_c$, for the CB case. Different values of $\Gamma$ are used 
($\Gamma=0.005,0.015,0.03,0.05$), and the other parameters are fixed to $\mathcal{P}=0.36$, 
$\Delta_0=1$, $\omega_{m,1}=1$. White regions correspond to unsynchronized regions.}
\end{figure}

The effect of dissipation on the synchronization threshold is quantified for CB and SB in Fig. \ref{figurethree}, for fixed 
  detuning, looking at the mechanical coupling $K_c^{thres}$ needed for the system to synchronize when increasing the damping rate $\Gamma$. 
   This mechanical coupling threshold $K_c^{thres}$ is defined as the minimum amount of mechanical coupling, $K_c$, needed for the oscillators to synchronize,
  and can be easily obtained as the synchronization of the system is signaled by a sharp transition of the synchronization indicator from small values 
  (for our parameters below $C_{max}\approx0.6$ )  to the values around unity that characterize synchronized oscillations.
 The damping strength does not influence significantly the synchronization threshold in SB case, in stark contrast with the CB case,  which is
 very sensitive to the damping strength and improving significantly when increasing $\Gamma$. The enlargement in the (anti-)synchronization area  for the CB case
 is also displayed in the full parameter plots (Fig. \ref{figurethree}b). 
 This clearly shows  how  the presence of an additional (dissipative) coupling term between the oscillators
 in the CB case triggers the emergence of spontaneous synchronization. In other words,
 both reactive and dissipative couplings contribute similarly in overcoming detuning effects in the context of synchronization.
 Furthermore CB specially favors anti-phase synchronization
 being the relative motion of the oscillators shielded by damping effects.
 
 An interesting effect of collective dissipation is that actually spontaneous synchronization can emerge even in absence of direct
 mechanical coupling (see Fig. \ref{figuretwo} for $\Delta\omega_m\lesssim 0.01$ and Fig. \ref{figurethree}a for $\Gamma\gtrsim 0.05$). 
 This is a specific signature of CB dissipation,
 being indeed not possible between uncoupled OMs with SB, and  shows the constructive role played by the 
 dissipative coupling due to CB even in absence of any other optical or mechanical coherent coupling.
The threshold for synchronization between decoupled oscillators ($K_c=0$) in the CB case is indeed set by the damping strength occurring for $\Gamma$ overcoming detuning effects
($\Delta \omega_m$) as nicely shown in the inset of  Fig. \ref{figurethree}a.
We also notice that, for $\Gamma$  small enough, synchronization is always in-phase (red points)
while when increasing it different phase (delays) can be favored at threshold (Fig. \ref{figurethree}).  In particular, in the CB case, 
the phase-locked difference does not always correspond to in-phase or anti-phase synchronization, as it also takes intermediate values  $\sim\pi/2$. This occurs both
with (Fig. \ref{figurethree}a) and without (inset in Fig. \ref{figurethree}a and Fig. \ref{figurethree}b) direct mechanical coupling, 
being more evident for larger dissipation rates (Fig. \ref{figurethree}a and b). The exact mechanism behind these $\pi/2$ values of the
locked phase-difference is an open question.

\section{Asymptotic entanglement and optomechanical cooling}
\label{sect4}

In this section we study the entanglement between mechanical modes in the asymptotic state of the dynamics. As explained in Section \ref{sect2} we set the system 
in a stable fixed point and we study the linear fluctuations around the mean 
state. The stationary state of the fluctuations is independent of the initial 
conditions and we characterize it obtaining numerically the stationary 
covariance matrix from equation (\ref{coveq}). From this covariance matrix we can compute entanglement between the 
different degrees of freedom \cite{50,51}, for which we use the logarithmic negativity, a valid 
entanglement monotone for mixed Gaussian states \cite{50,51}. Its definition for modes described by 
quadratures with commutation relations $[q_{1,2},p_{1,2}]=i$ is 
$E_N=\mathrm{max}\{0,-ln\,2\tilde{\nu}_-\}$, where $\tilde{\nu}_-$ is the smallest 
symplectic eigenvalue of the partially transposed covariance matrix of the 
subsystem \cite{51}.

One of the main results discussed below is the tight relation between the effective temperature of the mechanical modes and the amount of
asymptotic entanglement that is preserved between them: the linear regime
in which the system is operated basically consists of four coupled units which in the stationary state reach
a given effective temperature depending on the effectivity of the optomechanical cooling. It seems that this 
effective temperature regulates the entanglement behavior, which is that of a coupled thermal system.
Furthermore we find that the CB case has higher mechanical entanglement and is easier to cool down than 
the SB case, specially when increasing the relative mechanical coupling strength $K_c/\omega_m^2$.
We analyze next the dependence of cooling and entanglement on the different system parameters, showing
that this is a puzzle with several pieces.  At this point we remark that as in the following we are dealing with identical optomechanical oscillators,
quantities such as the occupation number of each of the mechanical oscillators ($n_{\mathrm{eff}}$), or the optical-mechanical entanglement between a mechanical mode
and its optical mode, are the same for the two optomechanical oscillators.

\begin{figure}[b!]
 \centering
 \includegraphics[width=1\columnwidth]{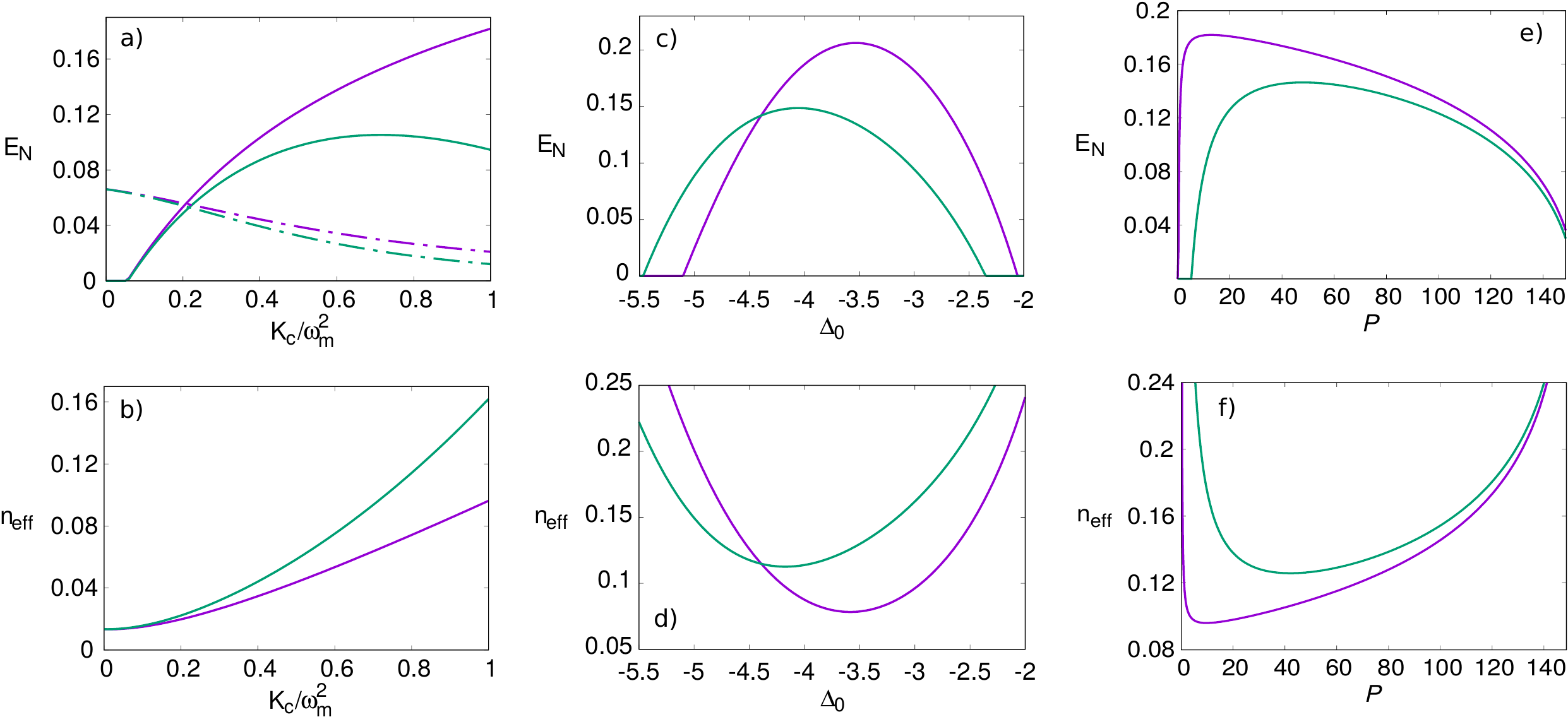}
 \caption{\label{figurefour} CB case: purple lines. SB case: green lines. Fixed 
parameters: $\omega_m=3$, $Q_m=\omega_m/\Gamma=10^5$ and $\hbar\omega_m/k_BT=0.1$ 
($n_{th}\approx9.5$). a) Asymptotic entanglement between the mechanical modes 
(solid lines) and between the mechanical modes and their respective optical 
modes (dashed-dotted lines)  as a function of the ratio $K_c/\omega_m^2$ 
This is an indicator of the 
frequency splitting of the normal modes of two coupled oscillators, i.e. 
$(\Omega_--\Omega_+)/\omega_m\approx K_c/\omega_m^2+O\big[(K_c/\omega_m^2)^2\big]$. 
b) Mechanical occupancy numbers. In a) and b) 
$\Delta_0=-\omega_m$ and $\mathcal{P}=12$. c,d) Asymptotic mechanical 
entanglement and mechanical occupancy numbers with $K_c/\omega_m^2=1$ and 
$\mathcal{P}=12$. e,f) Asymptotic mechanical entanglement and mechanical 
occupancy numbers with $K_c/\omega_m^2=1$ and $\Delta_0=-\omega_m$. }
\end{figure}

{\it Mechanical coupling $K_c$.-}  We first consider asymptotic entanglement when varying the mechanical
coupling between the
oscillators. A minimum  coupling is needed to
attain non-vanishing entanglement between the mechanical modes (solid
lines in Fig. \ref{figurefour}a), both for CB and SB. Indeed, only above some minimum coupling $K_c$  entanglement overcomes heating effects
as shown (through the mechanical occupancy number $n_{\mathrm{eff}}$) in Fig. \ref{figurefour}b. This is known to occur also
in the simplest configuration of a pair of coupled harmonic oscillators at finite temperature
\cite{anders,ent_fer}. We mention that actually for similar (effective) temperatures and similar mechanical coupling strengths we find that both harmonic and optomechanical
systems display similar amounts of entanglement.
Slightly above this threshold the build-up of entanglement
is similar in presence of CB or SB (Fig. \ref{figurefour}a), as also the heating of the mechanical components (Fig. \ref{figurefour}b),
but further increasing the coupling  differences arise
and entanglement worsen for SB. This effect is accompanied by a
stronger heating in the SB than in the CB case (Fig. \ref{figurefour}b). Overall, larger
entanglement, as well as better cooling  (smaller $n_{\mathrm{eff}}$), are found for
the CB case.

\begin{figure}[b!]
 \centering
 \includegraphics[width=0.9\columnwidth]{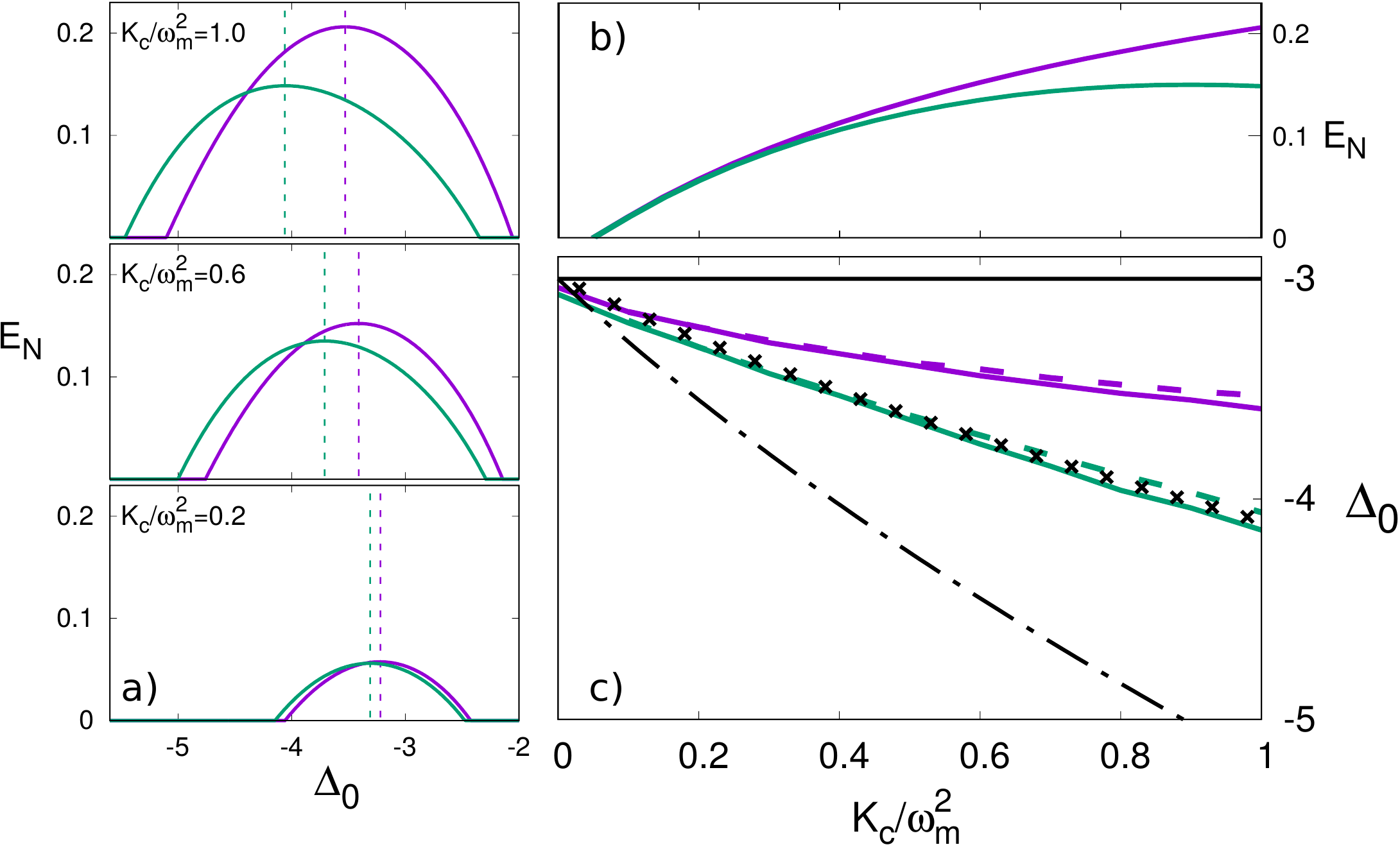}
 \caption{\label{figurefive} CB case in purple. SB case in green.  a) Entanglement for different values of $K_c/\omega_m^2=0.2,0.6,1$ as a function of $\Delta_0$ and with $\omega_m=3$, 
$Q_m=10^5$, $\hbar\omega_m/k_BT=0.1$ and $\mathcal{P}=12$.  Dashed lines at the maxima positions. b) Maximum 
entanglement for $\Delta_0$ as a function of $K_c/\omega_m^2$.  c) Solid colored lines: $\Delta_0$ maximizing optomechanical cooling as a function of $K_c/\omega_m^2$. 
 Dashed colored lines: $\Delta_0$ maximizing mechanical entanglement as a function of $K_c/\omega_m^2$.
The black solid line is $\Delta_0=-\Omega_+$, the black dashed-dotted is 
$\Delta_0=-\Omega_-$, and the black cross-line is $\Delta_0=-\overline{\Omega}$, 
defined in the main text.}
\end{figure}

From
these observations we infer that the mechanical coupling between the
oscillators influences their entanglement through two competing
mechanisms: on the one hand it is the coupling that entangles the two mechanical
oscillators, on the other $K_c$ modifies the effective temperature of the
oscillators. This is because the normal mode frequencies change with $K_c$ reducing their resonance with the optical modes
and therefore hindering their cooling (see discussion below and Fig. \ref{figurefive}). As the heating produced increasing $K_c$ is more
pronounced for
independent dissipation, best entanglement values are attained in the CB case.
This picture is confirmed when looking at ‘light-mirror’ entanglement:
$K_c$ is not expected to contribute significantly to the ‘light-mirror’ correlations in
each OM oscillator, while the effects of heating will still be present.
As we
see from figure \ref{figurefour}a (dashed-dotted lines)
the optical-mechanical entanglement is present for all couplings (no
threshold) and diminishes when increasing $K_c$. Both for CB and SB cases
the loss of entanglement is similar to the effective heating
(increasing effective temperature) in accordance with the above
considerations.

{\it Cavity-laser detuning $\Delta_0$.-} The detuning $\Delta_0$,  
set to values comparable to the mechanical frequencies, allows for quanta exchange between
the very disparate optical and mechanical modes. 
In this way the optical reservoir, effectively at $T=0$, acts as a heat sink for the mechanical 
degrees of freedom when the system operates in the red-sideband regime \cite{15,16}. Thus, resonance between optical and mechanical
degrees of freedom controls the effectiveness of the cooling process.  Looking at figures \ref{figurefour}c and \ref{figurefour}d we see that the entanglement
between oscillators and their effective temperature display a strong inverse behavior, i.e. as the effective temperature drops 
the entanglement grows and vice versa, and the maximum of entanglement is very close to the minimum of
temperature for both CB/SB cases. This behavior is found also for different temperatures and mechanical couplings.
Despite the fact that the optimal $\Delta_0$ values for cooling and entanglement
are slightly (less than $2\%$) shifted one with respect to the other, 
the strong anticorrelations between the two curves and the small size of these shifts, 
make us conclude
that the laser detuning $\Delta_0$ modifies the asymptotic mechanical entanglement mainly through the degree of optomechanical cooling.
We also note that the detuning for optimal cooling in both CB/SB cases is displaced with 
respect to known values for single OMs
$\Delta_0\approx-\omega_m$ \cite{15,16} (see discussion on optimal detuning). 
Finally we note that entanglement and cooling are again enhanced in the CB case, and that the optimal detunings for maximum entanglement are quite
different in the CB and SB case (the same applies to the detunings for optimal cooling). 
 
{\it Laser power $\mathcal{P}$.-}  Similarly to the case of $\Delta_0$, $\mathcal{P}$ modifies the mechanical entanglement  mainly by modifying 
the degree of optomechanical cooling  as it follows from the strong inverse behavior of figures \ref{figurefour}e and \ref{figurefour}f. Recall from the linear
quantum Langevin equations, that $\mathcal{P}$ has the role of a coupling strength between
the optical and mechanical quantum fluctuations. It is then to be expected that
optomechanical cooling improves with $\mathcal{P}$. However, when $\mathcal{P}$ becomes very 
high, it is known \cite{16,56} that optical and mechanical modes begin to hybridize and
heating from quantum backaction noise enters the 
scene. Thus at first higher $\mathcal{P}$ implies an enhancement of the optomechanical cooling effect,
 since this coupling parametrizes the strength of the optomechanical cooling rate \cite{15,16}. 
As $\mathcal{P}$ is further increased, the regime of strong coupling is reached and 
heating of the mechanical modes by radiation pressure becomes important. 
The competition between these two effects leads to a minimum of $n_{\mathrm{eff}}$ as a function of the optomechanical coupling strength,  as derived for a
single OMs in Refs. \cite{16,56} and observed in figure \ref{figurefour}f.
Finally, the CB case requires less power to achieve the same cooling as compared to the SB case.

{\it Optimal detuning.-}
 As we have observed in the discussion of figures \ref{figurefour}c and d, the asymptotic mechanical entanglement changes with $\Delta_0$ 
mainly because of the  
effectiveness of the optomechanical cooling process.
The detuning optimizing each process indeed takes very similar values as mentioned above.
In figure \ref{figurefive} we analyze in more detail the detunings $\Delta_0$ which are 
optimal for optomechanical cooling and mechanical entanglement as a function of the relative mechanical coupling 
strength $K_c/\omega_m^2$, and address some of the differences between the CB/SB cases that have been pointed out in the above discussions.
In Fig. \ref{figurefive}a we can see that as $K_c/\omega_m^2$ increases, the 
$\Delta_0$ maximizing entanglement is shifted towards more negative values. In figure \ref{figurefive}b
we have plotted the maximum entanglement for the optimal value of $\Delta_0$ as a 
function of $K_c/\omega_m^2$, and we see that it increases in both cases and it 
is larger for the CB case. We note that the increase with $K_c/\omega_m^2$ is similar to that in 
\ref{figurefour}a, but that here we compute the value at the optimal $\Delta_0$ for each $K_c/\omega_m^2$, $\textrm{max}_{\Delta_0}{ E_N(K_c/\omega_m^2)}$ , thus
increasing a bit the level of achieved entanglement.   Analogous results are obtained when studying the detuning for minimum effective temperature (not shown here).

 In figure \ref{figurefive}c we address the question of how the detunings that maximize entanglement or minimize the effective temperature vary with the relative mechanical coupling strength. In particular
we plot the $\Delta_0$ that optimizes cooling  (solid lines) and the one that optimizes entanglement (dashed lines) as a function of $K_c/\omega_m^2$. The solid black line 
corresponds to $\Delta_0=-\Omega_+$, the dashed-dotted line to 
$\Delta_0=-\Omega_-$, and the  cross-line to 
$\Delta_0=-\overline{\Omega}=-(\Omega_+ +\Omega_-)/2$. Where $\Omega_\pm$ are 
the frequencies of the normal modes of the isolated mechanical system 
$x_{\pm}=(x_1\pm x_2)/\sqrt{2}$ (i.e. without optomechanical coupling), and 
which take the values $\Omega_+=\omega_m$ and 
$\Omega_-=\sqrt{\omega_m^2+2K_c}$.  From this figure it is appreciated that the shift between the optimal detunings for cooling and entanglement increases slightly with the mechanical coupling, but it remains small
in all the range both for CB and SB.
We  also observe that in the SB case, the optimal strategy for cooling is to set 
($\Delta_0\approx-\overline{\Omega}$), and thus the minimum 
mechanical effective temperature is achieved when {\it both} mechanical normal modes (of the 
isolated system) are cooled at the same rate.  Notice that given the small difference between the detunings optimizing each quantity, the same strategy applies, to a good approximation, to obtain
the maximum entanglement.
On the other hand, in the CB case 
the optimal $\Delta_0$  for cooling and entanglement is shifted towards $-\Omega_+$,  consistent with the fact that
dissipation enters through this coordinate. Why is it then not the case that the optimal
detuning coincides with $\Omega_+$? Well, the common coordinate $x_+$ is a normal mode of the 
isolated mechanical system, but it is {\it not} of the full optomechanical system, so $x_+$ and $x_-$ are 
coupled through the optical degrees of freedom. This is why it is necessary to cool both coordinates,
although it is more important to cool the one that dissipates most ($x_+$). In fact, as we increase
$K_c/\omega_m^2$, the role of the optomechanical coupling is diminished, moving the optimal detuning 
towards $\Omega_+$. This also explains why the SB case is harder to cool than the CB one: in SB 
we need to address both normal modes $x_\pm$ with a single laser frequency, whereas for CB it is better to address only $x_+$.

\begin{figure}[H]
 \centering
 \includegraphics[width=1\columnwidth]{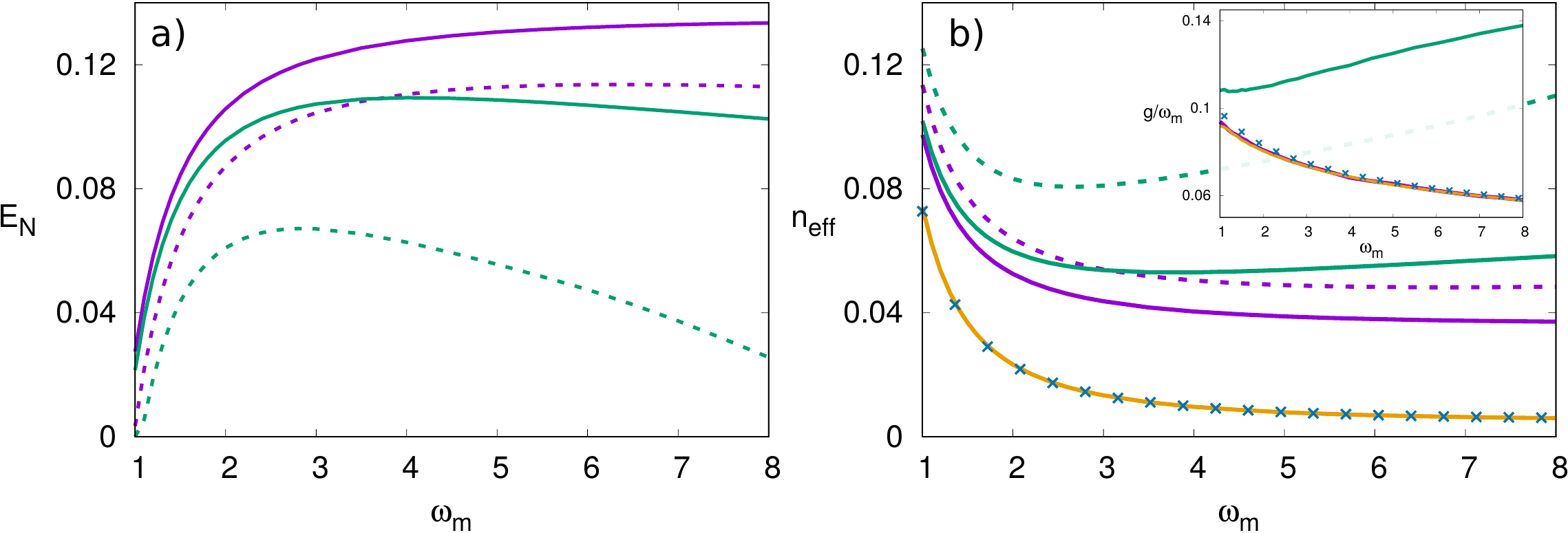}
 \caption{\label{figuresix} CB case: purple lines. SB case: green lines. Solid 
lines: $Q_m=10^5$, $\hbar\omega_m/k_BT=0.1$ ($n_{th}\approx9.5$). Dashed lines: 
$Q_m=2\cdot10^5$, $\hbar\omega_m/k_BT=0.01$ ($n_{th}\approx99.5$). Maximum 
entanglement (a) and minimum phonon number (b) optimized with respect to $\mathcal{P}$ (or equivalently with respect to $g/\omega_m$, defined in the main text) and varying 
$\omega_m$, with $K_c/\omega_m^2=0.5$, and $\Delta_0=-\omega_m$. Inset: optimal relative linear optomechanical coupling ($g/\omega_m$) at which the minimum phonon number is achieved, for the low
temperature case. In b) and in the inset the case of a single optomechanical oscillator has been plotted:
yellow solid lines have been obtained numerically, blue cross-lines correspond to the analytical expressions of Ref. \cite{56}.}
\end{figure}

{\it Resolved sideband.-}  In the following we explore mechanical entanglement (Fig. \ref{figuresix}a) 
and optomechanical cooling (Fig. \ref{figuresix}b)  as we go deeper into the resolved sideband
regime $\omega_m>1$, where the parameter $\omega_m$ measures how well the linewidth of the optical modes resolves the frequency of the mechanical ones. It is known that  
the efficiency of the cooling process of a single OM  increases initially as $\omega_m$
increases \cite{15,16,56} (recall that $\omega_m/\kappa \mapsto\omega_m $ in our dimensionless units, see Sect. \ref{sect2.2}), and thus lower phonon numbers can be achieved 
when the intracavity power is adjusted optimally \cite{56}. It is then interesting to analyze whether increasing $\omega_m$
can enhance cooling and entanglement in the CB/SB cases too. The maximal mechanical entanglement and the minimal mechanical occupancy number optimized in terms of
 $\mathcal{P}$ for each value of $\omega_m$  are displayed in figure \ref{figuresix}.

In figure \ref{figuresix}a (b), we can see how for the CB case the maximal entanglement (minimal occupancy number) increases (decreases) 
as $\omega_m$ is increased.  The SB case at first behaves as the CB case, but for larger $\omega_m$  maximal 
entanglement begins to diminish and the phonon number increases.  Notice that again the asymptotic mechanical entanglement and the
effective mechanical temperature are strongly anticorrelated
following an inverse behavior, in both CB/SB cases. The increase of the minimum effective temperature in the SB case is 
more evident in the high temperature case (dashed lines Fig. \ref{figuresix}a and b), and occurs even at optimal detuning $\Delta_0=-\overline{\Omega}$ (not shown here).
The difference between CB and SB cases can be explained as follows: for CB cooling the coordinate $x_+$ (by 
setting $\Delta_0=-\omega_m=-\Omega_+$) is almost optimal, and the more we resolve this frequency
with our optical modes the better. For SB however, we need to cool both $x_+$ and $x_-$,
but this has to be done with a single optical frequency: the result is that the better
we address $\Omega_+$, the worse we address $\Omega_-$ which gets heated up. Thus in a SB situation
it will be more appropriate to have one optical mode targeted at $\Delta_1=-\Omega_+$,
and another at $\Delta_2=-\Omega_-$, provided we can resolve the difference between $\Omega_+$ and
$\Omega_-$ which requires high mechanical couplings or very low $\kappa$.

 In order to point out further differences between the CB/SB cases it is also interesting to study the relative linear 
optomechanical coupling, {  $g/\omega_m$ (with $g=(\omega_c\tilde{x}_{zpf}/L_{om})\sqrt{0.5\langle{a^\dagger a}\rangle_{st}}$ \cite{1}, in
dimensional units)}, at which
optimal cooling is achieved (inset Fig. \ref{figuresix}b). From this figure we can observe that as we operate the system deeper in the resolved sideband 
regime (larger $\omega_m$) this optomechanical coupling
for optimal cooling diminishes in the CB case. This is so because the exchange between optical and mechanical modes
is more efficient, and it is the case both for one OM (yellow solid lines and blue cross-lines) \cite{56}, and for the CB case (where addressing $x_+$
matters most). In fact, if we increase it too much, we end up heating the mechanical modes due to
radiation pressure as shown in figure \ref{figurefour}f and in Refs. \cite{16,56}. Peculiarly, in the SB case, 
the optimal $g/\omega_m$ increases monotonically with $\omega_m$. This can be interpreted as follows: as before, 
the more we resolve, the worse we cool one of the coordinates $x_\pm$; the interesting thing is that this effect   is initially  
counteracted by increasing $g/\omega_m$ (see small values of $\omega_m$, Fig. \ref{figuresix}b), which increases the range of frequency that the mechanical mode is able to exchange 
energy with the optical modes. This is just the well-known fact that stronger couplings allow for resonance with further
detuned frequencies, which here counteracts the decreasing optical linewidth. However, as a side effect, 
radiation pressure heating also grows with the optomechanical coupling, and thus the maximum achievable entanglement (degree of cooling)
eventually decreases as shown in figure \ref{figuresix}a and b. 

 For a more quantitative discussion of cooling we simulate the case of a single OM (yellow solid lines and blue cross-lines, Fig. \ref{figuresix}b and inset),  
modeled by equations (\ref{lan1}) to (\ref{lan4}) for one of the two OMs in the SB case and with $K_c=0$. The case of a
single OM is studied in Ref. \cite{56}, where the optimization of the cooling process in terms of the linear optomechanical coupling is addressed. This is equivalent to what it is shown in figure
\ref{figuresix}b and in the inset, and we plot their results (Eq. (6) and below of \cite{56}) in blue cross-lines, using the parameters of the low temperature case.
We have also obtained the same quantities numerically (yellow solid lines)
finding good agreement between the analytical expressions of \cite{56} and the exact results. Comparing the single OM case and the CB case, we find that they follow very similar behaviors both for the
minimal occupancy number as well as for the optimal $g/\omega_m$ for cooling. In particular, lower phonon numbers are achieved in the single OM case, while the optimal $g/\omega_m$ for cooling takes very close
values in both cases. On the contrary, the behavior displayed in the SB case is qualitatively different as discussed above. The similitudes between the 
single OM case and the CB case reinforce the the overall idea that in the CB case we mainly need to address only one dissipative mode: $x_+$. 

\section{Discussion and conclusions}

Miniaturization and dense packing of optomechanical devices can give rise to
collective effects in dissipation \cite{31}, what we have called here common bath (CB), in 
contrast to the widely used separate baths (SB) dissipation. We have explored the consequences
of this form of dissipation in two interesting facets of these kind of nonlinear systems: 
classical synchronization and steady-state quantum correlations.
We have found that the parameter range where synchronization occurs is notably enlarged for CB,
and also that in the CB case more dissipation can be beneficial, leading to synchronization
even in the absence of mechanical coupling as both the dissipative coupling and the coherent one 
lead to synchronization between detuned mechanical elements. 

With respect to steady-state entanglement of
mechanical motion, we have found that the CB case requires less input laser power, because it
is more effectively used for cooling, yielding higher levels of entanglement. This can be understood
when considering that in this case it is possible to address only the $x_+$ mechanical eigenmode, which is the only one 
dissipating, whereas for SB both modes dissipate equally. This easy picture is quantitatively
shown to hold exploring different parameter regions. Furthermore we have shown that going to the resolved sideband regime, the CB case
gets monotonically better entanglement and cooling, whereas in SB it improves only up to a point, after
which it begins to worsen. This is also due to the inability to cool both eigenmodes with a perfectly
resolved sideband at only one of the eigenfrequencies, for which we also propose a two-tone cooling
scheme as solution.

As stated in the introduction, a collective dissipation channel is found to be the dominant one
in the optomechanical setup of references \cite{32,33}, consisting on the emission of elastic radiation by the common mechanical mode $x_+$, 
and thus being analogous to our CB model. Moreover the fact that this same mechanism is observed in other mechanical oscillators with a 
similar geometry \cite{36,37},
indicates that collective dissipation is relevant in closely clamped mechanical resonators. Despite the mechanical coupling strength
in this device \cite{32,33} is an order of magnitude smaller than the regime studied here, it is possible to
increase it up to the analyzed values by reducing the separation of the beams' clamping points and increasing the size of the overhang or common substrate, as shown in Ref. \cite{57}. 
Interestingly this last point suggests that strong mechanical coupling and collective dissipation might go hand in hand. On the other hand,
a complementary approach consists on inducing the mechanical coupling by strongly driving the system with off-resonant lasers \cite{38}. 

 In this work we have chosen to study the particular collective dissipation channel $x_+$ motivated by the numerous previous works studying this particular type of dissipation
emerging from microscopic physical models in
different contexts, starting from works on superradiance \cite{29}. Furthermore this collective dissipation has actually been reported in some mechanical platforms 
as discussed in the introduction
\cite{32,33,36,37}. However, other forms of collective dissipation are 
also possible: $x_-$ dissipation can arise in microscopic lattice models \cite{31}, and it has been reported in certain systems 
of trapped ions  \cite{ions}, while more complex forms of collective jumps are considered in \cite{zoller}. 
Moreover, quantum correlations and quantum synchronization for $x_-$ CB dissipation have been studied both for harmonic oscillators \cite{30,45}, for two level systems \cite{CB_ent},
and quantum Van der Pol oscillators \cite{extrasync2,extrasync4}. In general,
it would be interesting to study the robustness of our results also when both collective and individual dissipation channels are present, 
as this is probably 
the most common case in experiments.

Finally we discuss the experimental feasibility of the chosen parameters values, noticing that most of them have been reached in experimental platforms 
(see Ref. \cite{1} section IV for a review on the subject). Specifically, the dimensionless laser power is fixed to $\mathcal{P}=12$, which in terms of the 
more commonly quoted $g$, is $g/\omega_m\approx0.079$, and when varied takes values at most of $g/\omega_m\approx0.19$ in figure \ref{figuresix}, or 
$g/\omega_m\approx0.37$ when the dynamical instability is reached in figures \ref{figurefour}e,f. 
The values ranges in Fig. \ref{figurefour} have been achieved for instance in the experimental setups of references \cite{58,59,60},
 with the exception of the highest range values explored for $\mathcal{P}$ and $K_c/\omega_m^2$, for which previous considerations hold. 
 The resolved sideband regime has been reached in numerous optomechanical 
devices, from the GHz mechanical frequencies of optomechanical crystals \cite{42}, to MHz frequencies in microtoroidal and LC resonators where
values such as $\omega_m/\kappa\approx11$ \cite{59} or  $\omega_m/\kappa\approx63$ \cite{60} are reported. 
Moreover in optimized optomechanical crystals \cite{17,61}
bath phononic occupancies between 10 and 100 are reported for GHz mechanical modes, together with mechanical quality factors of the order of 
$Q_m\sim10^4-10^6$. Therefore, based on an analysis of actual 
state-of-the-art systems \cite{1}, this study represents a first step in establishing dynamical and quantum effects of collective dissipation 
expected to play a key role towards miniaturization of optomechanical devices.

\textit{Acknowledgments.-} This  work  has  been  supported  by  the  EU through the H2020 Project QuProCS (Grant Agreement 641277),  by  MINECO/AEI/FEDER  through  projects 
NoMaQ  FIS2014-60343-P,  QuStruct  FIS2015-66860-P and EPheQuCS FIS2016-78010-P. 

\appendix

\section{Derivation of the Langevin equations for the CB case}
\label{appA}

In this appendix we write down the main steps to derive the Langevin equations for the mechanical oscillators in the CB case.  Notice that in this section we are not using
the dimensionless variables and parameters introduced in Section \ref{sect2.2}. As usual we assume the cavities and
the oscillators to dissipate independently so that we can derive the Langevin equations separately, as if there was no optomechanical coupling \cite{1,Giov}. First we 
write down the mechanical Hamiltonian in the normal mode basis together with the bath Hamiltonians written in equation (\ref{hamCB}):

\begin{equation}
\mathcal{H}=\frac{1}{2}m(\Omega_+^2\hat{x}_+^2+\Omega_-^2\hat{x}_-^2)+\frac{1}{2m}\big(\hat{p}_+^2+\hat{p}_-^2 \big)+2\lambda\sum_{\alpha=1}
^{\infty}\hat{x}_+\hat{q}_\alpha+\hat{\mathcal{H}}^{CB}_{bath},\label{A.1}
\end{equation}

\noindent where we recall that we are assuming identical oscillators, and thus $\Omega_+=\omega_m$, $\Omega_-=\sqrt{\omega_m^2+2k/m}$. Making the rotating 
wave approximation in the system-bath coupling and writing the Hamiltonian using the creation and annihilation operators we arrive to:

\begin{equation}
\mathcal{H}=\hbar\Omega_+ \hat{b}_+^\dagger\hat{b}_++\hbar\Omega_- \hat{b}_-^\dagger\hat{b}_-+2\gamma\sum_{\alpha=1}
^{\infty}\hbar(\hat{b}_+^\dagger\hat{r}_\alpha+\hat{b}_+\hat{r}_\alpha^\dagger)+\hat{\mathcal{H}}^{CB}_{bath},\label{A.2}
\end{equation}

\noindent where $2\gamma$ is the CB dissipation rate which is twice the SB one $\gamma$. Now we can obtain the Langevin equations of motion for the modes $x_\pm$ following the 
standard input-ouput formalism \cite{62}:

\begin{equation}
\dot{\hat{b}}_+=-i\Omega_+\hat{b}_+-2\gamma\hat{b}_+ +\sqrt{4\gamma}\hat{b}_{in},\quad \dot{\hat{b}}_-=-i\Omega_-\hat{b}_-,\label{A.3}
\end{equation}

\begin{equation}
\dot{\hat{b}}^\dagger_+=i\Omega_+\hat{b}_+^\dagger-2\gamma\hat{b}_+^\dagger
+\sqrt{4\gamma}\hat{b}_{in}^\dagger,\quad \dot{\hat{b}}^\dagger_-=i\Omega_-\hat{b}_-^\dagger,\label{A.4}
\end{equation} 

\noindent where the equations for the mode $x_-$ are just the Heisenberg equations of motion, since this mode is not coupled to a thermal bath. In the Markovian limit the 
zero mean Gaussian noise terms are characterized by the following correlations \cite{22}:

\begin{equation}
\langle  \hat{b}_{in}^\dagger(t)\hat{b}_{in}(t')\rangle=n_{th}\delta(t-t') ,\quad \langle  \hat{b}_{in}(t)\hat{b}^\dagger_{in}(t')\rangle=(n_{th}+1)\delta(t-t'),\label{A.5}
\end{equation}

\noindent where $n_{th}$ is the thermal occupancy  of the phonon bath defined previously. From these equations we can obtain the equations for the position and momentum 
operators:

\begin{equation}
\dot{\hat{x}}_+=\frac{\hat{p}_+}{m}-2\gamma\hat{x}_++\sqrt{4\gamma}\,\tilde{x}_{zpf}\big[\frac{\hat{b}_{in}+\hat{b}_{in}^\dagger}{\sqrt{2}}\big],\label{A.6} 
\quad \dot{\hat{x}}_-=\frac{\hat{p}_-}{m},
\end{equation}

\begin{equation}
\dot{\hat{p}}_+=-m\Omega_+^2\hat{x}_+-2\gamma\hat{p}_++\sqrt{4\gamma}\,m\Omega_+\tilde{x}_{zpf}\big[i\frac{\hat{b}^\dagger_{in}-\hat{b}_{in}}{\sqrt{2}}\big], 
\quad \quad \dot{\hat{p}}_-=-m\Omega_-^2\hat{x}_-.\label{A.7}
\end{equation}

Finally we move to the coupled oscillators picture, where $\hat{x}_{\pm}=(\hat{x}_1\pm\hat{x}_2)/\sqrt{2}$ and $\hat{p}_{\pm}=(\hat{p}_1\pm\hat{p}_2)/\sqrt{2}$, 
obtaining:

\begin{equation}
\dot{\hat{x}}_j=\frac{\hat{p}_j}{m}-\gamma(\hat{x}_1+\hat{x}_2)+\tilde{x}_{zpf}\hat{x}_{in}, \quad j=1,2\label{A.8}
\end{equation}

\begin{equation}
\dot{\hat{p}}_j=-m\omega_m^2\hat{x}_j-\gamma(\hat{p}_1+\hat{p}_2)+(-1)^j k(\hat{x}_1-\hat{x}_2)+
m\omega_m\tilde{x}_{zpf}\hat{p}_{in},\label{A.9}
\end{equation}

\noindent where we have defined $\hat{x}_{in}=\sqrt{\gamma}(\hat{b}_{in}+\hat{b}_{in}^\dagger)$ and $\hat{p}_{in}=i\sqrt{\gamma}(\hat{b}_{in}^\dagger-\hat{b}_{in})$. From these 
equations we can obtain equations (\ref{lan3}) and (\ref{lan4}) by coupling the oscillators to the cavities and applying the nondimensionalization/rescaling procedure described
in Section \ref{sect2.2}. Furthermore, note that as the noise terms are the same for both oscillators there appear the cross-correlations defined in equation (\ref{crossnoise}).

\end{document}